\documentclass{article}
\usepackage{amsmath,amsfonts,amssymb,mathrsfs}
\usepackage[all]{xy}
\usepackage{graphicx}
\usepackage{amsthm}
\usepackage{psfrag}
\usepackage{color}
\usepackage{authblk}

\begin{document}

\title{Dynamical features of the MAPK cascade}
\author[1]{Juliette Hell}
\author[2]{ Alan D. Rendall}
\affil[1]{Freie Universit\"at Berlin, { \tt jhell@zedat.fu-berlin.de}.}
\affil[2]{Johannes Gutenberg-Universit\"at Mainz, { \tt rendall@uni-mainz.de}.}
\date{}

\maketitle

\abstract{The MAP kinase cascade is an important signal transduction system in 
molecular biology for which a lot of mathematical modelling has been done. 
This paper surveys what has been proved mathematically about the
qualitative properties of solutions of the ordinary differential equations 
arising as models for this biological system. It focusses, in particular, on
the issues of multistability and the existence of sustained oscillations.
It also gives a concise introduction to the mathematical techniques used
in this context, bifurcation theory and geometric singular perturbation
theory, as they relate to these specific examples. In addition further 
directions are presented in which the applications of these techniques 
could be extended in the future.}

\section{Introduction}

An important process in cell biology is the transmission of information 
by signalling networks from the cell membrane to the nucleus, where it can
influence transcription. This provides the cell with a possibility of 
reacting to its environment. A common module in many signalling networks
is the mitogen activated protein kinase cascade (MAPK cascade). It is the
subject of what follows. The MAPK cascade is a pattern of chemical
reactions which is widespread in eukaryotes \cite{widmann99}. The individual 
proteins which make up the cascade differ between different organisms and 
between different examples within a given organism but what is common 
is a certain architecture. The cascade consists of three parts which we will 
call layers. Each layer is a phosphorylation cycle or, as it is sometimes 
called, a multiple futile cycle \cite{wangsontag08a}.

A multiple futile cycle consists of a protein $X$ which can be phosphorylated
by a kinase $E$ at $n$ sites. The resulting phosphoproteins can be 
dephosphorylated by a phosphatase $F$. The cases of interest for the MAPK 
cascade are $n=1$ and $n=2$. The following sketch shows the MAPK cascade with 
three layers. Each plain arrow marked with an enzyme represents an enzymatic 
reaction, i.e 
$\xymatrix@C=0.5em{
Y\ar[rrr]^{G}&&&Z
}$
stands for the chemical reactions $Y+G \rightleftarrows YG \rightarrow Z+G$. The 
dotted arrows between a phosphorylated protein and an enzyme in the next layer 
stand for equality, i.e. 
$\xymatrix@C=0.5em{
Z\ar@{.>}[rrr]&&&G
}$ means $Z=G$. 
The first layer of the cascade is a simple phosphorylation, i.e. $n=1$. The 
next layers are double phosphorylations, i.e. $n=2$. 
The species $X_i$ are the proteins in the cascade, or in more detail
$X_1=MAPKKK$ (MAP kinase kinase kinase), $X_2=MAPKK$ (MAP kinase kinase), 
$X_3=MAPK$ (MAP kinase). 
\begin{equation}\label{cascade}
\xymatrix@C=0.5em{
 X_1\ar@/^1pc/[rr]^{E_1}&&X_1P\ar@/^1pc/[ll]^{F_1}   \ar@{.>}[dl]  \ar@{.>}[dr]&&&&\\
 &&&&&&\\
 X_2\ar@/^2pc/[rr]^{E_2} &&X_2P\ar@/^2pc/[rr]^{E_2} \ar@/^1pc/[ll]^{F_2}  &&X_2PP\ar@/^1pc/[ll]^{F_2} \ar@{.>}[dl]  \ar@{.>}[dr] &&\\
 &&&&&&\\
 &&X_3\ar@/^2pc/[rr]^{E_3}&&X_3P\ar@/^2pc/[rr]^{E_3} \ar@/^1pc/[ll]^{F_3}&&X_3PP\ar@/^1pc/[ll]^{F_3}  
}
\end{equation}
It is important that even when there is more 
than one site where the protein can be phosphorylated there is only one
kinase and one phosphatase. Thus different phospho-forms of the protein may
compete for binding to one of the enzymes. A MAPK cascade consists 
of three layers, each of which is a multiple futile cycle with a 
different protein. 
The connection between the layers is that the maximally phosphorylated 
forms of the protein in the first and second layers are the kinases which 
catalyse the phosphorylations in the second and third layers, respectively.
The most extensively studied example in mammals is that where the proteins
in the three layers are Raf, MEK and ERK. 

The subject of what follows is mathematical modelling of the MAPK cascade.
It turns out that this system has a rich dynamics which needs mathematical 
modelling for its understanding. Pioneering work in studying this
question was done by Huang and Ferrell \cite{huang96}. In that paper the 
authors presented both theoretical and experimental results and compared them.
Their experiments where done in cell extracts from {\it Xenopus} oocytes.
On the theoretical side they wrote down a system of ordinary differential 
equations describing the time evolution of the concentrations of the 
substances involved in the MAPK cascade and simulated these equations
numerically. The results of the simulations reproduced important 
qualitative features of the experimental data.     

In order to model the reactions taking place it is necessary to make 
assumptions about the kinetics. In many enzymatic reactions the 
concentrations of the enzymes are much less than those of the corresponding 
substrates. This cannot necessarily be assumed in the case of the MAPK 
cascade. In particular there are substances, for example the doubly
phosphorylated form of MEK, which are the substrate for one reaction 
and the enzyme for others. For this reason the model in \cite{huang96}
uses a Michaelis-Menten scheme for each reaction with a substrate, an
enzyme and a substrate-enzyme complex without making an assumption of small enzyme concentration. The elementary reactions involved
are assumed to obey mass action kinetics. There are seven conservation laws 
for the total amounts of the three substrates and the four enzymes which are 
not also substrates (one kinase, $E_1$ and three phosphatases, $F_i$, $i=1,2,3$). It is assumed that 
the phosphorylation and dephosphorylation are distributive and sequential.
In other words in any one encounter of a substrate with an enzyme only
one (de-)phosphorylation takes place, after which the enzyme is released.
To add or remove more than one phosphate group more than one encounter is
necessary. The phosphorylations take place in a particular order and the 
dephosphorylations in the reverse order. These assumptions are implemented
in the model of \cite{huang96}. They have also frequently been adopted
in other literature concerned with the modelling of this system and we will
call them the standard assumptions. The authors of \cite{huang96} mention 
that they also did simulations for cases where one or more of the reactions is 
processive (i.e. more than one phosphate group is added or removed during
one encounter). The standard assumptions may not be correct in all 
biological examples but they are a convenient starting point for modelling 
which can later be modified if necessary.

In real biological systems the MAPK cascade is part of a larger signalling 
network and cannot be seen in isolation. Nevertheless, one can hope to
obtain insights by first understanding the isolated cascade and later 
combining it with other reactions. Similarly it can be helpful to 
approach an understanding of the cascade itself by studying its component 
parts, the multiple futile cycles.

The most frequent approaches to these modelling questions in the literature
use simulations and heuristic considerations. An alternative possibility,
which is the central theme of this paper, is to prove mathematical theorems
about certain aspects of the dynamics with the aim of obtaining 
insights complementary to those coming from the numerical procedures. The 
number of mathematically rigorous results on this subject known up to now is 
rather limited. The aim of this paper is to survey the results of this type 
which are available and to outline perspectives of how they might be extended. 
At the same time it gives an introduction to some of the techniques which are 
useful in this kind of approach. The description starts from the simplest 
models and proceeds to more complicated ones. We discuss successively the 
simple futile cycle, the dual futile cycle and the full cascade. The 
description also proceeds from simpler dynamical features to more complicated
ones, from multistationarity to multistability and then to sustained 
oscillations. After this core material has been treated further directions
are explored. What happens when the basic cascade is embedded in feedback 
loops? What happens in systems with other phosphorylation schemes?

\section{The simple futile cycle}

In this section we look at the case $n=1$ of the multiple futile cycle, in 
other words we isolate the first layer of the cascade \eqref{cascade}.
We omit the index 1 of the chemical species for clarity in this section, since 
the other layers will play no role here. 
Modelling this system in a way strictly analogous to that applied to the 
MAPK cascade in \cite{huang96} leads to a system of six equations for the 
substrates $X$ and $XP$, the enzymes $E$, $F$ and the substrate-enzyme 
complexes $X\cdot E$ and $XP\cdot F$. There are three conserved quantities,
which are the the total amounts of the enzymes and the substrate,
$E_{\rm tot}=[E]+[X\cdot E]$, $F_{\rm tot}=[F]+[XP\cdot F]$ and 
$X_{\rm tot}=[X]+[XP]+[X\cdot E]+[XP\cdot F]$. 
where here and in the following  $[Z]$ denotes the concentration of the 
species $Z$.
These can be used to eliminate three of
the equations if desired.
When the evolution is modelled by mass-action kinetics, these manipulations 
can be done explicitly.

Suppose we have a system $\dot x_i=f_i(x)$, 
$x=(x_1, \ldots, x_n)\in \mathbb{R}^n$ representing the dynamics of a 
chemical reaction network. The $x_i(t)$ are the concentrations of the substances
involved as functions of time and the dot denotes the time rate of change. A 
stationary solution (or steady state) is one which satisfies for all 
$i\in \{ 1, \ldots, n\}$, $x_i(t)\equiv x_{i,0}$ 
for some fixed concentrations $x_0=(x_{1,0}, \ldots, x_{n,0})$. Thus it 
satisfies $\dot x=0$ or 
equivalently $f(x_0)=0$. A solution $x(t)$ is said to converge to the steady 
state $x_0$ if $\lim_{t\to\infty} x(t)=x_0$. This is an idealization of the 
situation where an experimental system settles down to a steady state on 
a sufficiently long time scale. For instance in the experiments of 
\cite{huang96} the system was found to approach a steady state after 100
minutes. Corresponding behaviour was found in the simulations. There is no 
reason why a chemical system should behave in this simple way for all 
initial data. The results of \cite{huang96} indicate that it does so for
the data considered there. Even if for a particular system all solutions 
converge to a steady state, it may be that there exist more than one
steady state for fixed values of the total amounts of the substances
involved. In the language of chemical reaction network theory \cite{feinberg80},
there may be more than one steady state in one stoichiometric compatibility 
class. This is the phenomenon of multistability. It is important for biological 
processes such as cell differentiation.
  
In the case of the simple futile cycle it was proved in \cite{angeli06} that 
there is always exactly one steady state for fixed values of the total amounts 
$E_{\rm tot}$, $F_{\rm tot}$ and $X_{\rm tot}$ and that all other solutions 
converge to that steady state. The steady state is globally asymptotically 
stable and bistability is ruled out in this case. We cannot enter into the
details of the proof of this result here but it is appropriate to mention
some of the key ideas involved. Suppose that the system $\dot x_i=f_i(x)$
has the property that $\partial f_j/\partial x_i>0$ for all $i\ne j$ and all
$x$. Then the system is called monotone. If a system does not satisfy this 
property we may try to make it do so by reversing the signs of some of the 
variables. In other words we replace the variables $x_i$ by 
$y_i=\epsilon_i x_i$, where each $\epsilon_i$ is plus or minus one. In 
general a system is called monotone if there is a mathematical transformation 
of this kind which makes all partial derivatives of the right hand sides of 
the equations with $i\ne j$ positive. There is a graphical criterion to decide 
whether this is possible. Define a graph which has a vertex for each variable 
$x_i$ and which has an oriented edge connecting node $i$ to node $j$ if 
$\partial f_j/\partial x_i\ne 0$. Label each edge with the sign of the 
corresponding derivative. Alternatively we can use the convention that a 
positive sign is represented by a normal arrow while a negative sign is 
represented by a blunt-headed arrow. This object is called the species 
 graph. For example the  species 
 graph of the simple futile cycle is 
the following.
\begin{equation}
\xymatrix{
& XP \ar@{<->}[dd]^+ \ar@{<->}[dl]_{-} \ar@(ul,l)_{-}& XE \ar[l]_+ \ar@{<->}[rd]^+ \ar@{<->}[dd]_+ \ar@(ur,r)^{-} & \\
F \ar@(ul,l)_{-} \ar@{<->}[dr]_{+}&&& E \ar@{<->}[ld]^{-} \ar@(ur,r)^{-}\\
& XPF\ar@(dl,l)^{-} \ar[r]_+ & X \ar@(dr,r)_{-} &
}
\end{equation} 
Since the use of terms concerning feedback loops is not always consistent between different sources in the literature we specify the terminology which we will use. 
A feedback loop is a sequence of arrows which starts at one node and ends at 
the same node, i.e. a cycle. It is called a positive or negative 
feedback loop according to whether the number of edges with a negative sign 
it contains is even or odd. More precisely this object may be called a 
directed feedback loop, while the corresponding definition where the orientation
of the edges is ignored is called an undirected feedback loop. Suppose we have 
a system of ordinary differential equations for which the sign of the 
derivative $\partial f_j/\partial x_i$ is independent of $x$ for each fixed 
$i$ and $j$. Then it can be proved that the system is monotone if and only if 
the species 
 graph contains no negative undirected feedback loops of length
greater than one. In the case 
of the simple futile cycle the species 
 graph contains at least one negative 
feedback loop.  This is true both for the full six-dimensional system and for 
the three-dimensional system obtained by eliminating the concentrations of 
$E$, $F$ and $XP$ using the conservation laws. However, it was shown in 
\cite{angeli06} 
that there is a different type of transformation which makes this system 
monotone. In this transformation the concentrations are replaced as variables 
by the extents of the reactions. The resulting monotone system has additional 
good properties and this allows the property of convergence to a unique steady 
state to be concluded for the original system.

Many chemical systems have interesting limiting cases obtained by letting
certain combinations of reaction constants tend to zero. This can lead to
a significant reduction in the number of variables in the system and 
make analytical investigations simpler. Under suitable circumstances 
solutions of the limiting system approximate solutions of the original
system in a certain parameter regime. This can be illustrated by the case
of the simple futile cycle. For the species $E,F,X\cdot E,XP\cdot F$
involving the enzymes $E$, $F$, alone or in a complex, scale the 
concentrations with a parameter $\epsilon>0$ while the remaining concentrations $[X]$ and $[XP]$ are not rescaled. In other words, if 
$y=([E],[F],[X\cdot E],[XP\cdot F])$ is the vector of their concentrations, 
define $\tilde y=\epsilon y$, where $\epsilon$ is a positive constant. 
If $x=([X],[XP])$ is the vector of the remaining concentrations, the original system with
mass-action kinetics is of the form
\begin{equation}\label{maoriginal}
\begin{cases}
\dot x = f(x,y),\\
\dot y = g(x,y),
\end{cases}
\end{equation}
where $f,g$ are linear in each of the concentrations, i.e. entries of the 
vectors $x,y$. 
The smallness of $\epsilon$ corresponds to the fact that the amount of enzymes 
is small compared to the amount of the other species. Define
a new time coordinate by $\tau=\epsilon t$ and let a prime denote the 
derivative with respect to time $\tau$. 
The time $\tau$ is called the slow time scale because its velocity 
$\frac{d\tau}{dt}=\epsilon$ is small. 
This gives rise to a system of the general 
form 
\begin{equation} \label{slow}
\begin{cases}
x'=f(x,\tilde y),\\
\epsilon \tilde{y}'=g(x,\tilde y).
\end{cases}
\end{equation}
In the limit $\epsilon\to 0$, and dropping the tildes, the second 
equation $\epsilon y'=g(x,y)$  changes from being an ordinary differential 
equation to being
an algebraic equation $0=g(x,y)$. Under favourable conditions this equation 
can be
solved for $y$ in terms of $x$ and the result $y=h(x)$ substituted into the 
first 
equation to give an equation for $x$ alone, 
\begin{equation}\label{MMred}
x'=f(x,h(x)).
\end{equation}
 The result is a system with fewer
unknowns. The degeneration to an algebraic equation means that the limit is 
singular. It may be asked whether the solutions themselves nevertheless behave 
in a regular way in the limit and indeed this is the case under certain 
conditions. The appropriate mathematical techniques for studying this are 
known as geometric singular perturbation theory (GSPT) as introduced by Fenichel, see \cite{fenichel79}. This theory will be discussed 
in more detail below. The particular type of limit just exhibited for the 
simple futile cycle is sometimes called a Michaelis-Menten limit. In this case 
it leads to a two-dimensional system. The conservation law for the total 
amount of substrate survives in the limit in a simplified form and can be used to reduce the system 
further to a single equation. Let the reaction constants for
complex formation, complex dissociation and product formation be denoted
by $k_i$, $d_i$ and $a_i$ respectively, with $i=1$ corresponding to 
phosphorylation and $i=2$ to dephosphorylation. 
\begin{equation}
\xymatrix@C=0.5em{
X+E\ar@/^1pc/[rrr]^{k_1}&&&XE\ar@/^1pc/[lll]^{d_1} \ar[rrr]^{a_1}&&&XP+E\\
\\
XP+F\ar@/^1pc/[rrr]^{k_2}&&&XP\cdot F \ar@/^1pc/[lll]^{d_2} \ar[rrr]^{a_2}&&&X+F
}
\end{equation} 
The Michaelis constants are
defined as usual by $K_{m,i}=\frac{d_i+k_i}{a_i}$. Then the equation can be
written as 
\begin{equation}
\frac{d}{dt}[XP]=-\frac{k_2F_{\rm tot}(X_{\rm tot}-[XP])}{K_{m,2}+X_{\rm tot}-[XP]}
+\frac{k_1E_{\rm tot}[XP]}{K_{m,1}+[XP]},
\end{equation}
After this reduction it is possible to get an explicit formula for the unique 
steady state by solving a quadratic equation in the variable 
$[XP]$ \cite{goldbeter81}. For 
$\epsilon$ small the total amounts of the enzymes are small compared to the 
total amount of substrate and this is sometimes described by saying that the 
enzymes are close to saturation. Varying one of the parameters in the system 
and monitoring the concentration of phosphorylated protein gives a response 
function. The main concern of \cite{goldbeter81} is the form of this function, 
which corresponds to the property of ultrasensitivity: A small change in the parameter leads to a large change in the value of the response function. This property is 
quantitative rather than qualitative and not obviously amenable to the 
application of the analytical techniques to be discussed in this paper. It is 
an interesting question, whether these techniques can be extended so as to 
give more information about quantitative properties. This will not be discussed 
further here except to mention that ultrasensitivity was also the central 
feature of interest in \cite{huang96}, where the response of the concentration 
of maximally phosphorylated ERK as a function of that of the first kinase in 
the cascade was investigated.

\section{The dual futile cycle}

This section is concerned with the case $n=2$ of the multiple futile cycle.
The second layer of the MAPK cascade \eqref{cascade} is an example
of a dual futile cycle. 
We omit the index indicating the layer in this section, since we consider a 
single layer of the cascade. 
The basic system with mass action kinetics can be found, for instance, in
\cite{wangsontag08a}. It is possible to do a Michaelis-Menten reduction
in a similar way to that done in the last section. 
We recall that this consists in scaling the concentration $Z$  of the species 
containing one of the two enzymes $E$, $F$, alone or in a complex, via the 
transformation $\tilde{y}=\epsilon y$ of their concentration vector $y$, as 
well as the time by $\tau=\epsilon t$. The limit $\epsilon\rightarrow 0$ can
be reduced to a lower dimensional ODE by solving an algebraic equation.
This leads, after using all conservation laws to eliminate as many variables 
as possible, to a two-dimensional system, which will be called the MM system
(for Michaelis-Menten). More details on 
this can be found in \cite{hell14}. The system can be written in the form
\begin{eqnarray}
&&\frac{d}{dt}[X]=-\frac{k_1K_{m,1}^{-1}E_{\rm tot}[X]}
{1+K_{m,1}^{-1}[XP]+K_{m,3}^{-1}[XPP]}
+\frac{k_2K_{m,1}^{-2}F_{\rm tot}[XP]}
{1+K_{m,2}^{-1}[XP]+K_{m,4}^{-1}[XPP]}.\\
&&\frac{d}{dt}[XPP]=\frac{k_3K_{m,3}^{-1}E_{\rm tot}[XP]}
{1+K_{m,1}^{-1}[XP]+K_{m,3}^{-1}[XPP]}
-\frac{k_4K_{m,4}^{-1}F_{\rm tot}[XP]}
{1+K_{m,2}^{-1}[XP]+K_{m,4}^{-1}[XPP]}.
\end{eqnarray}
When using the conservation law for $X_{\rm tot}$ we have the choice, which of
the three concentrations $[X]$, $[XP]$ or $[XPP]$ to eliminate. Here the 
equation for $[XP]$ has been discarded and on the right hand side of the 
equations $[XP]$ should be regarded as an abbreviation for 
$X_{\rm tot}-[X]-[XPP]$. In this case the MM system is monotone, as defined in 
the previous section.
Since it is two-dimensional this implies that all solutions converge to
steady states. Moreover it can be shown using GSPT that for $\epsilon$
small but non-zero almost all initial data give rise to solutions which 
converge to steady states \cite{wangsontag08b}. 
For general $\epsilon$ it was not known until very recently whether the 
corresponding 
statement was true. In \cite{errami15} the authors used computer-assisted 
methods to find periodic solutions of this system, indicating that the 
statement is false for general $\epsilon$. They do not only use 
dynamical simulations but also use computer algebra to help implement a 
theoretical approach to finding Hopf bifurcations. They do not obtain evidence 
for the existence of stable periodic solutions, so that it would be 
consistent with their findings if periodic solutions, while present, were
only relevant for exceptional initial conditions. Despite the results of
\cite{errami15} the global
behaviour of solutions of the dual futile cycle is much less well understood
than that in the case of the simple futile cycle.

What has been proved is that multistationarity (existence of more than 
one steady state) occurs in the dual futile cycle for certain values of the 
parameters \cite{wangsontag08a}. In fact it is known that there exist up to
three steady states for given values of the total amounts and that there are
never more than three. The proof of the existence  
result can be split into two steps. In the first step the equations for steady
states are partly solved explicitly. This leaves a system of two equations
for two unknowns. The second step involves taking a limit of these equations
as a parameter $\epsilon$ tends to zero. This limit is essentially the 
Michaelis-Menten limit discussed above. Since we are dealing with steady
states the factor $\epsilon$ in the equation for $y$ plays no role and the
limit is regular. In the limit a single equation for one variable is obtained
and this is relatively easy to analyse. It is possible to find cases where 
there are three solutions of the equation $F(x)=0$ for steady states of
the equation with $\epsilon=0$, all of which satisfy
$dF/dx\ne 0$. This implies, using the implicit function theorem, that 
corresponding solutions exist for $\epsilon$ small but positive. This argument 
gives no information on the important issue of stability of these solutions. A
steady state $x_0$ will only be observed in practise if it is stable. This 
means that if a solution starts close enough to $x_0$ it will stay close to
$x_0$ for all future times. For example if the linearization at the 
equilibrium $x_0$ has only eigenvalues with strictly negative real parts, then 
the equilibrium $x_0$ is  stable. 

On the level of simulations multistationarity was already observed for the 
system with mass action kinetics in \cite{markevich04} and it was found that
two of the three steady states are stable. Simulations in 
\cite{ortega06} indicated that these features are already present in the
MM system. On the other hand until recently there was no mathematical proof
of bistability for the dual futile cycle. A strategy suggested by what has 
been said up to know for obtaining such a proof is to first prove
bistability for the MM system and then use GSPT to conclude the 
corresponding result for the mass action system. This strategy was carried out
in \cite{hell14} and in that paper we proved bistability. We now sketch the 
main lines of the proof. 

\begin{figure}[h]
\includegraphics[width=1\textwidth]{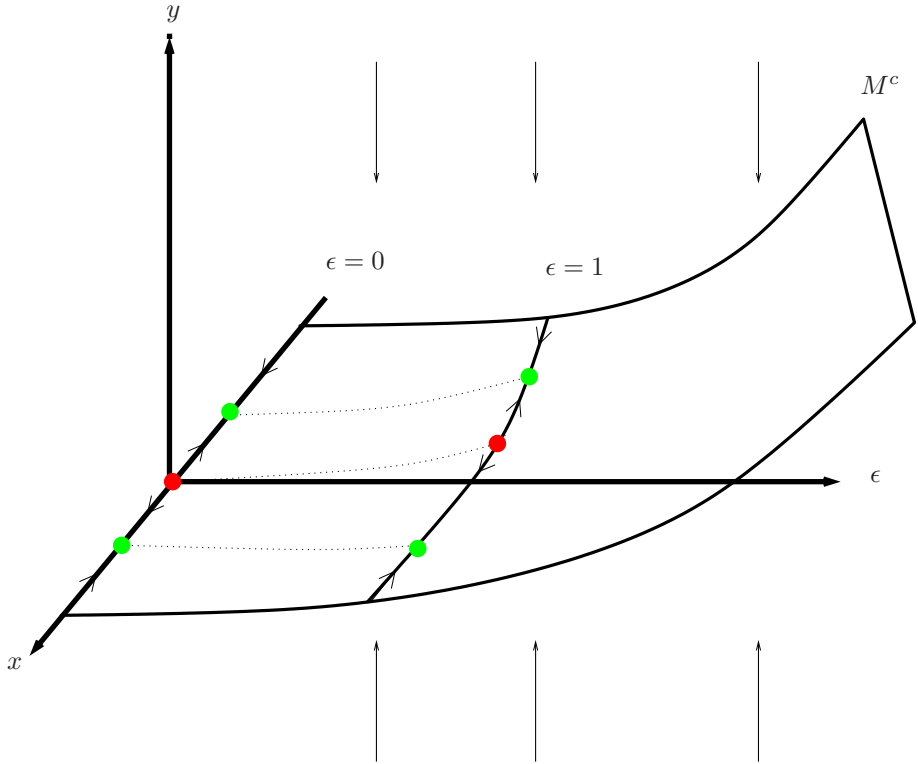}
\caption{ \label{gspthyp} Sketch of the phase portrait of the extended system with a curve of equilibria $\{y=h(x)\equiv 0 \}$. The one dimensional dynamics of the MM-reduced system consists of one unstable equilibrium connected via heteroclinic orbits to two stable equilibria. This dynamics persists on the invariant $\epsilon$-leaves for small $\epsilon$.}
\end{figure}

In the previous section, we saw that the MM reduction is the singular limit as 
$\epsilon\rightarrow 0$ of the system on the slow time scale \eqref{slow}, 
where $g(x,\tilde y)=0  \Leftrightarrow \tilde y=h(x)$. 
Now consider the fast time scale, i.e. the same equations expressed in the 
original time $t$, augmented by the trivial evolution of the parameter 
$\epsilon$. This system is called the extended system. 
\begin{equation}\label{fastext}
\begin{cases}
\dot{x}=\epsilon f(x,\tilde y, )\\
 \dot{\tilde{y}}=g(x,\tilde y, )\\
\dot{\epsilon}=0
\end{cases}
\end{equation}
The curve $\{ (x,\tilde{y}=h(x),\epsilon=0)\}$ is a curve of equilibria in 
the  extended system. The linearization of the extended system \eqref{fastext} 
at such an equilibrium admits two zero eigenvalues with the corresponding 
eigenvector  being $(0,0,1)$ pointing in direction of $\epsilon$ 
orthogonally to the $(x,y)$-plane, and a second vector in the $(x,y)$-plane 
tangent to the curve $\{(x,h(x))\}$. If the linearization $D_y g (x,h(x),0)$ 
admits only eigenvalues with strictly negative real parts, then the center 
eigenspace (i.e. the eigenspace spanned by the eigenvectors associated to 
purely imaginary eigenvalues) is exactly two dimensional. 
The eigenspace is tangential to a manifold called the center manifold: see \cite{vdbw} for details about center manifold theory. 
The (local) center 
manifold $M^c  (x,h(x),0)$ of such an equilibrium is an invariant manifold 
containing all  bounded solutions sufficiently near the equilibrium.  The 
center manifold $M^c$ can be written as a graph of a function $\Psi$ over the 
center eigenspace:
\begin{equation}
M^c(x,h(x),0)=\{ (x,\Psi(x,\epsilon), \epsilon), \epsilon \text{ small} \},
\end{equation}
with $\Psi(x,0)=h(x)$. The manifold $M^c$ is tangent to the center eigenspace  
of the extended system at each equilibrium $(x,h(x),0)$. 
Because of the equation $\dot \epsilon =0$, the invariant 2-dimensional center 
manifold $M^c$ is foliated by invariant 1-dimensional leaves 
$\epsilon=constant$. If the leaf at $\epsilon=0$ consists of hyperbolic 
equilibria (i.e. the linearization there admits no purely imaginary 
eigenvalues) connected by heteroclinic orbits as depicted in figure \ref{gspthyp}, 
then this dynamics is preserved in the leaves at $\epsilon$ small. 
Furthermore, the linearization theorem of Shoshitaishvili (see \cite{Kusnetsov}, Theorem 5.4) tells us that, if 
$D_y g(x,h(x))$ has only stable eigenvalues, the center manifold $M^c$ is attracting. 
Hence hyperbolic equilibria that are stable in the MM-reduced system 
\eqref{MMred} give rise to stable equilibria in the extended system. Undoing 
the scaling $\tilde y = \epsilon y$ provides us with stable equilibria for the 
original mass-action system \eqref{maoriginal}.  
\psfrag{input}{input}
\psfrag{output}{output}
\psfrag{lambda}{$\lambda$}
\psfrag{l=1}{$\epsilon=1$}
\psfrag{l=0}{$\epsilon=0$}
\psfrag{x}{$x$}
\psfrag{y}{$y$}
\psfrag{epsilon}{$\epsilon$}
\psfrag{Mc}{$M^c$}

In order to apply GSPT in this context the main property to be checked 
concerns the eigenvalues of the matrix $A=D_y g(x,h(x))$.
By this we mean the matrix of partial derivatives of the function $g$ with 
respect to the variables $y$ at a point $(x,h(x))$. The condition to be checked for the center 
manifold $M^c$ to be attracting is that the real
parts of these eigenvalues are negative. Fortunately in this case the 
calculation can be reduced to one for the eigenvalues of two by two matrices,
which is relatively simple.  

The remaining part of the argument is to prove bistability for the MM system. 
When this has been done and if the stable steady states are hyperbolic (i.e. 
the linearization of the system at those points have no eigenvalues with zero 
real parts) then GSPT tells us that these steady states continue to exist 
and be stable for $\epsilon$ small but non-zero. For more details we refer to 
\cite{hell14}. Bistability for the MM system is proved using bifurcation 
theory and now some more information will be given concerning that technique. 
Consider a system of ordinary differential equations $\dot x=f(x,\alpha)$ and 
a steady state $(x_0,\alpha_0)$, i.e. $f(x_0,\alpha_0)=0$. Here $\alpha$ 
denotes one or more parameters. The linearization of the system at 
$(x_0,\alpha_0)$ is the matrix of partial derivatives $A=D_xf (x_0,\alpha_0)$. 
If no eigenvalue of $A$ has zero real part the equilibrium $x_0$ is said to be 
hyperbolic. Then for $\alpha$ close to $\alpha_0$ there is a unique solution of 
$f(x_*(\alpha),\alpha)=0$ with $x_*$ close to $x_0$ by the implicit function 
theorem. The stability properties of the solution are preserved, the dynamics 
nearby is the same as the dynamics of the linearized system by the Hartman-Grobman 
Theorem (see \cite{guckho}, Theorem 1.3.1). For instance if $x_0$ is 
stable then $x_*(\alpha)$ is stable. If, on the other hand, $A$ has an
eigenvalue whose real part is zero then $(x_0,\alpha_0)$ is said to be a 
bifurcation point and the qualitative dynamics of the system may change
at that parameter value. For instance as the parameter is varied one steady 
state may split into several. In other words, new branches of equilibria may 
come into being at the critical parameter value $\alpha_0$.  Identifying a 
suitable bifurcation is a way of proving 
that several steady states exist for certain parameter values.

\begin{figure}[h]
\includegraphics[width=1\textwidth]{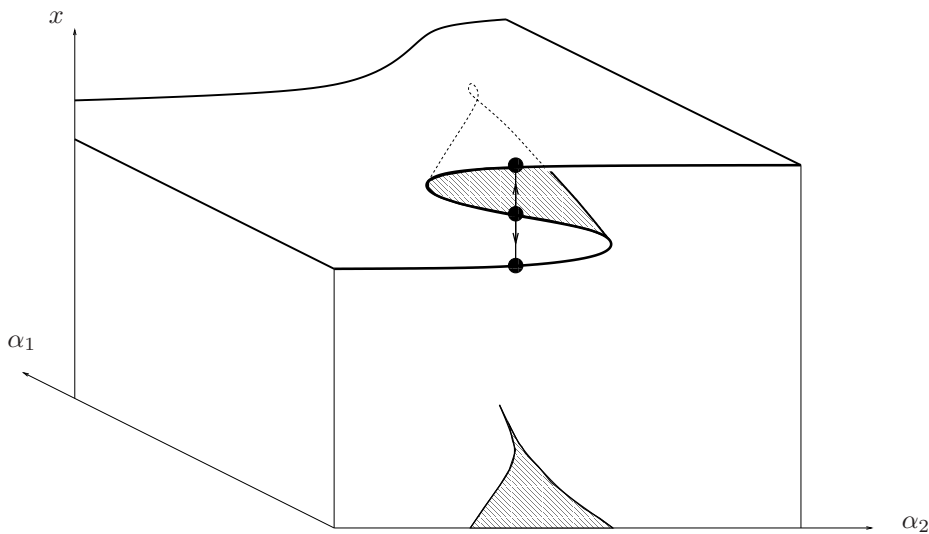}
\caption{ \label{cuspfig} Bifurcation diagram for a generic cusp bifurcation: the unstable branch of the surface of equilibria is shaded, as well as  the region in the parameter plane where two stable and one unstable  equilibria coexist.}
\end{figure}

For example, consider a system $\dot x=f(x,\alpha)$ with one unknown $x$ depending on two parameters, $\alpha=(\alpha_1,\alpha_2)$.
Denote by a prime the partial derivative of a function of $x$ and $\alpha$ with respect to $x$. 
 Suppose
that $f(0,0)=0$, $f'(0,0)=0$, $f''(0,0)=0$ and $f'''(0,0)\ne 0$ and that an
additional quantity depending on derivatives with respect to the parameters
is non-zero. The eigenvalue of the linearization $A$ that crosses zero depends 
on the two parameters. 
In this example there is only one eigenvalue of the Jacobian, which is $f'(0,0)=0$.
When the above mentioned quantity is non-zero, it 
guarantees that the crossing happens at a non-zero velocity with respect to 
the parameters and transversally to the imaginary axis. This is called a 
transversality condition.  See \cite{Kusnetsov} for details.  
These are the defining properties of what is called a generic
cusp bifurcation. 
There is a surface of equilibria over the two dimensional parameter space 
which develops a fold at $(0,0)$. In a cusp region of the parameter space, 
three equilibria coexist - two stable ones and an unstable one. See Figure 
\ref{cuspfig}. 
\psfrag{1mu}{$\alpha_1$}
\psfrag{2mu}{$\alpha_2$}
\psfrag{x}{$x$} 
Then there are parameter values near zero for which the 
system has three steady states close to zero. 
The case of relevance for the examples considered in this paper is that where
$f'''(0,0)<0$ and from now on we will only discuss that case. There two of
the steady states close to zero are stable and one unstable.
 Now 
suppose there are several variables $x_i$ and that at the point $(0,0)$ the
derivative $A=D_xf$ has a zero eigenvalue of multiplicity one. The kernel
of $A$ is of dimension one. The qualitative behaviour of solutions near the 
steady state is determined by the restriction of the dynamics to a 
curve, the  one dimensional centre manifold, which is tangent to the kernel of $A$ and 
invariant under the flow. While the local stable and unstable manifolds 
contain all initial conditions for solutions that converge exponentially to 
the equilbrium in forward or backward time direction respectively, the center 
manifold contains local bounded dynamics that depends heavily on the nonlinear 
terms. In this way 
the general case may be reduced to the one-dimensional case already discussed 
when the kernel of $A$ is of dimension one. 
In fact the dynamics in the stable and unstable directions corresponding to 
eigenvalues with nonzero real parts do not change. 
These are the main techniques used 
in the proof of bistability in \cite{hell14}.

It is clear that in order to linearize about a steady state we first have to
have that steady state. It is not too difficult to find steady states since
there are many parameters in the problem which can be varied. In \cite{hell14}
steady states were considered for which the concentrations of $X$ and $XPP$
are equal, since this simplifies the algebra. Moreover it was assumed that
all Michaelis constants have a common value $K$. In this case the relation
\begin{equation}
\left(\frac{E_{\rm tot}}{F_{\rm tot}}\right)^2
=\frac{k_2k_4}{k_1k_3}
\end{equation}
holds. A bifurcation point was found under the restriction that
$q^2=(k_1k_4)/(k_2k_3)<1$. The bifurcation occurs when 
$KX_{\rm tot}=\frac{2+q}{1-q}$. 

It is important to note that here Michaelis-Menten reduction was carried out
for the whole system and not for the two phosphorylation steps separately.
The latter alternative leads to a different set of equations. It was used
in \cite{kholodenko00}, where the effect of embedding a MAPK cascade in a
negative feedback loop was investigated. Consider, for instance, the dual
futile cycle which is the third layer of the cascade in \cite{kholodenko00}
and the equation for the unphosphorylated protein. It is of the form
\begin{equation}
\frac{d}{dt}[X]=-\frac{k_7E_{\rm tot}[X]}{K_7+[X]}
+\frac{k_7F_{\rm tot}[XP]}{K_{10}+[XP]}
\end{equation}
which is clearly different from the corresponding equation in the MM system 
introduced above, even when all Michaelis constants are taken to be equal 
(which is done in the simulations of \cite{kholodenko00}).  


The results just discussed only give limited information on the region of
parameter space in which bistability occurs. In contrast, in \cite{conradi14} 
rigorous quantitative results on multistationarity were proved. Rather general
conditions on the reaction constants were exhibited for which there
are  one or three steady states.

\begin{figure}[h]
\includegraphics[width=0.8\textwidth]{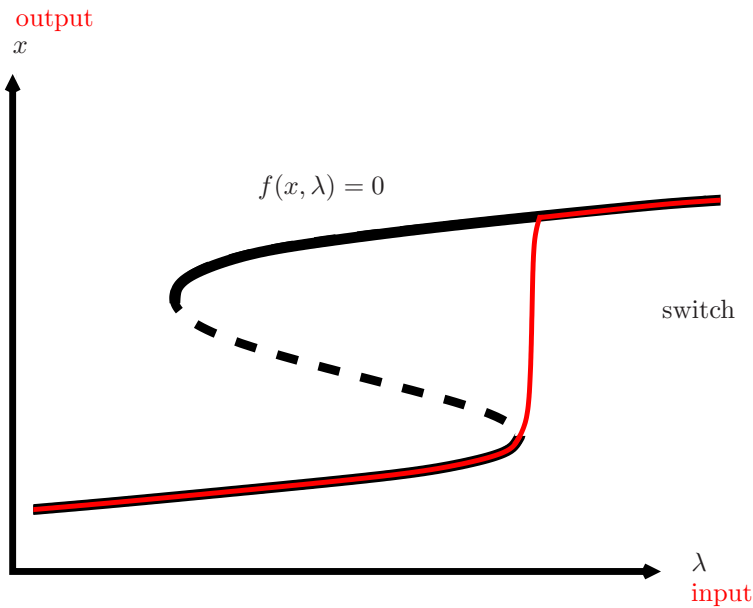}
\caption{ \label{switch} The \textcolor{red}{input-output relation} follows the lower stable equilibria until it becomes unstable through a fold (also called saddle-node) bifurcation, then jumps to the upper stable  branch. Hence such an input-output relation is a good switch. }
\end{figure}

Bistability is important for the property of being a `good switch':  
consider our system as an input-output relation where
the input consists of the values of the conserved quantities and the output 
consists of the equilibrium reached by the system after a certain amount of 
time. For certain values of the input, the stable equilibrium reached by the 
system is unique and, say, on the lower part of the folded surface of 
equilibria. When the  input enters the cusp region, the output remains on the 
lower part of the fold until the cusp region is left: at this point, the 
output switches to the upper part of the folded surface. 
In fact, the same phenomenon of switching is present when the parameter 
(input) is one dimensional and there exists a S-shaped curve of equilibria. 
See figure \ref{switch}.

\psfrag{input}{\textcolor{red}{input}}
\psfrag{output}{\textcolor{red}{output}}
\psfrag{lambda}{$\lambda$}
\psfrag{switch}{switch}
\psfrag{f=0}{$f(x,\lambda)=0$}
Sometimes when modelling phosphorylation systems the enzyme concentrations are not included explicitly and instead mass action kinetics is used for the substrates alone. If this is done for the cycle with two phosphorylations, or indeed for a cycle with any number of phosphorylations, a dynamical system is obtained which has a unique stationary solution to which all other solutions converge. This is because it can be shown that the system is a weakly reversible system of deficiency zero and the Deficiency Zero Theorem of chemical reaction network theory (see \cite{feinberg80}) can be applied. This type of argument was used to prove corresponding results for the kinetic proofreading model of T cell activation in \cite{sontag01} and for the multiple phospho- rylation of the transcription factor NFAT in \cite{rendall12}. These systems only describe small parts of the network involved in T cell activation, which also contains a MAPK cas- cade. A comprehensive model of this phenomenon presented in \cite{alboger05} is too large to be accessible by direct analytical investigation. Interestingly a very much simplified version of this model introduced in \cite{fvsabv} reproduces some of the key features of T cell activation such as specificity, speed and sensitivity. In the simplified model the MAPK cascade is represented by a simple response function.
\section{The MAPK cascade}

The starting point of this section is the model of the MAPK cascade introduced
in \cite{huang96}. Simulations in \cite{qiao07} revealed the presence of 
bistability and sustained oscillations in this model. Mathematically the 
latter correspond to periodic solutions, i.e. solutions which satisfy
$x(t+T)=x(t)$ for some time interval $T$ but are not steady states. In 
\cite{qiao07} similar results were obtained for the truncated cascade 
consisting of just the first two layers. Michaelis-Menten reduction for the 
MAPK cascade is made difficult by the fact that there is not a clear division 
between substrates and enzymes. This issue was studied in \cite{ventura08}. A 
further development of these ideas in \cite{ventura13} indicated that periodic 
solutions already occur in the Michaelis-Menten limit. It was shown 
in \cite{hell14} that a small modification of these ideas allows a
Michaelis-Menten limit of the equations for the truncated cascade to be
defined which is well-behaved in the sense of GSPT. 
Since the truncated system contains a species, $X_1P$, which is a product in 
the first layer and an enzyme in the second layer, the $\epsilon$-scaling of 
the MM-reduction has to be carried out using two different powers of 
$\epsilon$. We first define  a new variable $\overline{X_1}$ replacing $[X_1P]$  as follows.
\begin{equation*}
\overline{X_1}:= [X_1P]+[X_2\cdot X_1P]+ [X_2P\cdot X_1P],
\end{equation*}
The concentration vector is split into three vectors $v_0,v_1,v_2$. The 
vector $v_2$ is the vector of concentrations of species containing the enzymes 
of the first layer of the cascade, alone or in a complex, i.e. 
$v_2=([E_1], [F_1], [X_1\cdot E_1], [X_1P\cdot F_1])$. This vector is rescaled 
by $\tilde{v_2}=\epsilon^2 v_2$
The vector $v_1$is the vector of concentrations of species containing
$X_1$ or the enzymes $X_1P=E_2$ and $F_2$ of the second layer of the cascade, 
alone or in a complex, i.e. 
$v_1=(\overline{X_1},[X_1],[X_1P], [E_2], [X_2\cdot X_1P]
, [X_2P\cdot X_1P],[X_2P\cdot F_2],[X_2PP\cdot F_2] )$. This vector is rescaled 
by $\tilde{v_1}=\epsilon v_1$.
Finally, the vector $v_0$ contains the concentrations of the  remaining 
species, i.e.  $x=([X_2], [X_2P], [X_2PP])$ and  is not rescaled.
Furthermore the reaction constants of the first layer are also rescaled by 
$\epsilon$. A slow time variable $\tau= \epsilon t$ is introduced: the time 
derivative w.r.t. $t$ is denoted by an upper dot while the time derivative 
w.r.t. $\tau$ is denoted by $'$. Using conserved quantities to reduce the 
dimension of the concentration vectors and the new variable $\overline{X_1}$, we 
get a system of the form 
\begin{equation}
\begin{cases}
x' = f(x,y,z),\\
\epsilon y'= g(x,y,z),\\
\end{cases}
\end{equation}
where $x=(\overline{X_1}, [X_2], [X_2PP])$ and $y=([X_1\cdot E_1],[X_1P\cdot F_1],[X_2\cdot X_1P],[X_2P\cdot X_1P],[X_2P\cdot F_2],[X_2PP\cdot F_2])$
The limit $\epsilon\rightarrow 0$ of this system allows a MM-reduction that is 
well-behaved in terms of GSPT. For more details see \cite{hell14}, 
\cite{hell15}. This result was extended
to the full cascade in \cite{hell15}. 

The facts just listed indicate that the strategy used to prove 
bistability in the dual futile cycle might also be used to prove the 
existence of sustained oscillations in the truncated MAPK cascade, i.e. 
layers 1 and 2 of cascade \eqref{cascade}. Here the relevant 
type of bifurcation is a Hopf bifurcation where the linearization at an equilibrium admits  a pair of 
imaginary eigenvalues for a critical  parameter. By a classical theorem of Hopf, if under variation of
a parameter a pair of complex conjugate eigenvalues of the linearization passes through the imaginary
axis (away from zero) with non-zero velocity then there exist periodic
solutions for at least some parameter values.
See \cite{Kusnetsov}, \cite{guckho} for details. 
 In \cite{hell15} it was proved
that Hopf bifurcations occur in the MM system for the MAPK cascade. As a
consequence periodic solutions occur.
%
If an additional genericity condition 
(hyperbolicity of the periodic orbit) were satisfied then it could further be 
concluded using GSPT that periodic solutions also occur in the mass action 
system for the truncated MAPK cascade. In fact it has not yet been possible to 
prove hyperbolicity. Instead
 it was proved that the Hopf bifurcation itself 
can be lifted to the mass action system and this then gives the existence of 
periodic solutions of that system. 
Unfortunately these arguments give no 
information on the stability of the periodic solutions involved. It is 
interesting to look at this situation in the light of results on feedback 
loops. It has been proved that a system can only admit a stable periodic 
solution if it includes a directed negative feedback loop \cite{angeli09}, 
\cite{richard10}. It can easily
checked directly that this condition holds in the case of the MM system for
the truncated MAPK cascade.


The results for the truncated cascade imply analogous results for the full
cascade by another application of GSPT. What must be shown is that the 
truncated cascade can be represented as a limit of the full cascade which
is well-behaved in the sense of GSPT. Consider the MM system for the full
cascade. Let $Z$ be the concentration of a protein in the first or second 
layer and define a new variable by $\tilde Z=\epsilon^{-1}Z$. Let 
$c_i$ be any of the rate constants in the third layer and define 
$\tilde c_i=\epsilon c_i$. The transformed system has a limit for 
$\epsilon\to 0$ which is well-behaved in the sense of GSPT and the limiting 
system is the MM system for the truncated cascade. Thus in the context of
the MM system the Hopf bifurcation can be lifted from the truncated to the 
full cascade. It can then be further lifted from the MM system for the full
cascade to the mass action system.   

In \cite{prabakaran14} an {\it in vitro} model of the MAPK cascade was 
introduced. The substances involved are modified in such a way that certain
features of the reaction network are modified. In the first layer Raf is
constitutively active which means that for modelling this layer can be 
ignored. In the third layer ERK can only be phosphorylated once, on tyrosine 
and not on threonine.
(In the wild type system MEK has the unusual property of being a dual
specificity kinase which can phosphorylate both threonine and tyrosine.)
 This leads to a cascade with two layers where the 
first has two phosphorylation steps and the second only one:
\begin{equation}
\xymatrix{
\mathrm{Raf} \ar@{.>}[dr] \ar@{.>}[drrr] & &&&\\
&&&&\\
X_2 \ar@/^2pc/[rr]^{E_2} & & X_2P \ar@/^2pc/[rr]^{E_2} \ar@/^1pc/[ll]^{F_2}&& X_2PP \ar@{.>}[dl] \ar@/^1pc/[ll]^{F_2}\\
&&&&\\
&&X_3 \ar@/^2pc/[rr]^{E_3} & & X_3P \ar@/^1pc/[ll]^{F_3} 
}
\end{equation}
Here $X_3=ERK$.
This was 
modelled mathematically in a certain way in \cite{prabakaran14} and it was
proved that for that system of equations there is a unique steady state and
all other solutions converge to it. 
If, on the other hand, the system is
modelled in direct analogy to what was done in \cite{huang96} a system is obtained
which might potentially admit Hopf bifurcations and hence periodic solutions.
It was written down in \cite{hell15} but attempts to prove the existence of Hopf
bifurcations using the methods applied to the truncated MAPK cascade have
not succeeded.
Simulations done in \cite{zumsande10} indicate that there may be chaotic 
behaviour in the MAPK cascade. The approach of the authors was via 
(numerical) bifurcation theory. They discovered the presence of 
fold-Hopf bifurcations (where the linearization of the system at the 
bifurcation point has one zero and a pair of non-zero imaginary eigenvalues)
and Hopf-Hopf bifurcations (where the linearization of the system at the 
bifurcation point has two pairs of non-zero imaginary eigenvalues). 
See \cite{Kusnetsov} for more details on these types of bifurcations. 
Bifurcations of these types are often associated with chaos. Simulations
for initial data close to the bifurcation points gave pictures consistent
with the presence of chaotic behaviour.      

In \cite{qiao07} a heuristic explanation for the existence of oscillations in the MAPK cas- cade was given, involving the embedding of a bistable system in a negative feedback loop. This point of view played no role in the proof of the existence of periodic so- lutions using bifurcation theory which has just been discussed. It could, however, in principle lead to an alternative proof of that result which could also provide in- formation on the stability of the periodic solution. A corresponding strategy, which makes use of the Conley index (see \cite{conley78}), has been developed and applied to a system related to that considered in \cite{kholodenko00}. More details can be found in \cite{afs04}, \cite{gs07} and \cite{gedeon10}.

\section{Embedding the cascade in feedback loops}

Given that modelling predicts oscillatory behaviour in the MAPK cascade it
is of great interest to try to observe it experimentally. This has been 
done in \cite{shankaran09}. Oscillations were found in the concentration
of activated (i.e. doubly phosphorylated) ERK which had a period of about 
fifteen minutes and lasted up to ten hours. This effect was monitored by
observing the translocation of ERK tagged with GFP between the cytosol and
the nucleus. This is relevant because activated ERK is imported into the 
nucleus and inactive ERK is exported into the cytosol. Mathematical models
presented in \cite{shankaran09} are more complicated than the basic model of
\cite{huang96} in several ways. 

 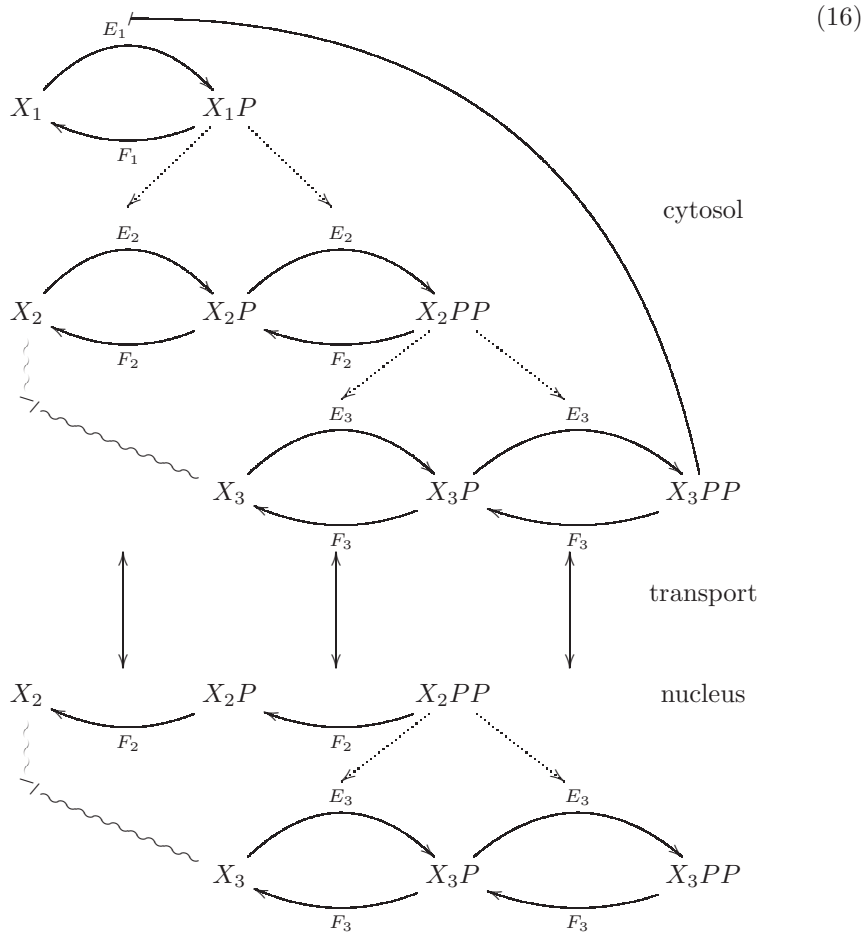
\begin{figure}[h]
\begin{equation}
\xymatrix{
&&&&&&\\
X_1\ar@/^2pc/[rr]^{E_1\quad}&&X_1P\ar@/^1pc/[ll]^{F_1}   \ar@{.>}[dl]  \ar@{.>}[dr]&&&&\\
&&&&&&\text{cytosol}\\
X_2\ar@/^2pc/[rr]^{E_2} \ar@{~/}[d]&&X_2P\ar@/^2pc/[rr]^{E_2} \ar@/^1pc/[ll]^{F_2}  &&X_2PP\ar@/^1pc/[ll]^{F_2} \ar@{.>}[dl]  \ar@{.>}[dr] &&\\
&&&&&&\\
&\ar@{}[dd]^(.25){}="a"^(.9){}="b" \ar@{<->} "a";"b"
&X_3\ar@/^2pc/[rr]^{E_3} \ar@{~/}[llu]&\ar@{}[dd]^(.25){}="a"^(.9){}="b" \ar@{<->} "a";"b"&X_3P\ar@/^2pc/[rr]^{E_3} \ar@/^1pc/[ll]^{F_3}&\ar@{}[dd]^(.25){}="a"^(.9){}="b" \ar@{<->} "a";"b"&X_3PP\ar@/^1pc/[ll]^{F_3}\ar@{-/}@/_5pc/[uuuuulllll]\\ 
&&&&&&\text{transport}\\
X_2\ar@{~/}[d]&&X_2P \ar@/^1pc/[ll]^{F_2}  &&X_2PP\ar@/^1pc/[ll]^{F_2} \ar@{.>}[dl]  \ar@{.>}[dr] &&\text{nucleus}\\
&&&&&&\\
&&X_3\ar@/^2pc/[rr]^{E_3} \ar@{~/}[llu]&&X_3P\ar@/^2pc/[rr]^{E_3} \ar@/^1pc/[ll]^{F_3}&&X_3PP\ar@/^1pc/[ll]^{F_3}   
}
\end{equation}
\caption{ \label{cascadeinloop} Two-compartement model with a full MAPK cascade in the cytosol, a lower half of the MAPK cascade in the nucleus, transport of $X_3PP$ from the cytosol to the nucleus and transport of $X_3$ from the nucleus to the cytosol.}
\end{figure}
One is that a two-compartment model is used so 
that transport between cytosol and nucleus is included. 
This network is sketched in Figure \ref{cascadeinloop} below, where Raf $= X_1$,  MEK $= X_2$, ERK $ = X_3$.
A second is that the 
fact is included that ERK and MEK which are not fully phosphorylated can 
bind to each other ($\xymatrix{\ar@{~/}[r]& &\ar@{~/}[l]}$ in Figure \ref{cascadeinloop}). This protects ERK from phosphorylation by MEK and thus 
represents a kind of negative feedback. Note that the tendency of this type
of feedback to encourage bistability was already pointed out in \cite{legewie07}.
A third way is that a negative feedback due the repression of SOS by ERK is 
included ($\xymatrix{\ar@{-/}[r]& }$ in Figure \ref{cascadeinloop}). This is important since SOS influences the rate of phosphorylation 
of Raf. This last effect is modelled in a simple phenomenological manner. It 
was discovered that certain aspects of the experimental data do not fit the 
mathematical model with an isolated cascade. In particular this concerns the 
facts that the oscillations are found for a large range of total amounts of 
ERK and that the period of the oscillations is found to depend only weakly on 
the total amount of EGF, the substance being used to stimulate the cascade. It 
is assumed that the phosphorylation of Raf is proportional to the amount of EGF.
Incorporating the negative feedback loop via SOS in the model allows the
experimental results to be reproduced.

Yet another type of negative feedback which may lead to oscillations arises
via the competition of a substrate of ERK with a phosphatase being reduced
by increasing degradation of the activated substrate \cite{liu11}. In this 
paper the authors present both mass action and reduced models and find
periodic solutions in simulations.

\section{Alternative phosphorylation mechanisms}

In this paper we have concentrated on distributive phosphorylation. If this
is partly replaced by processive (but still sequential) phosphorylation then 
this often results in simpler dynamics. For instance it was proved in 
\cite{conradi05} that when phosphorylation, dephosphorylation or both are
replaced in the dual futile cycle by their processive versions,  then there is a unique steady state for fixed 
values of the total amounts. This follows from the deficiency one algorithm
of chemical reaction network theory. The result was generalized to the
analogue of the multiple futile cycle with strictly processive and 
sequential phosphorylation in \cite{conradi15}. In addition it was proved 
that all other solutions converge to the steady state. Note that 
other types of (partially) processive phosphorylation may also be 
considered \cite{gunawardena07}.

One of the most important roles of oscillations in biology is that they can
act as clocks, for instance those defining circadian rhythms. Many of these
clocks are dependendent on translation but a clock has been found in 
cyanobacteria which is not. It uses only phosphorylation states of the 
proteins KaiA, KaiB and KaiC. This has been demonstrated by reconstructing
the clock {\it in vitro} \cite{nakajima05}. In KaiC the phosphorylation is
cyclic rather than sequential. In other words the first of two sites to be
phosphorylated is also the first to be dephosphorylated. This motivated the
study of oscillations in dual phosphorylation models more general than the
usual dual futile cycle \cite{jolley12}. In that paper the case of 
unordered phosphorylation was considered, i.e. that where the two 
(de-)phosphorylations may take place in any order. Simulations indicate that
this is sufficient to produce sustained oscillations.

While the MAPK cascade is a type of phosphorylation system of central
importance in eukaryotic signalling pathways, signalling pathways in
prokaryotes more often use a different type of phosphorylation system
known as a two-component system \cite{stock00}. These are uncommon in
eukaryotes and unknown in mammals. The central mechanism is as follows.
The two components are proteins generically called HK and RR. The protein
HK is a histidine kinase which, under appropriate conditions, phosphorylates
itself on a histidine. RR, the response regulator, catalyses the transfer of
the phosphate group from the histidine of HK to an aspartate in RR. In
this way the RR is activated. HK can also dephosphorylate RR.

What is the advantage of the process of phosphorylation of RR taking
place in two steps rather than directly? It has been suggested that
the motivation is that the two-step process leads to absolute
concentration robustness \cite{shinar10}. This means that the output signal is
independent of the total concentrations of the enzymes, so that the
system achieves independence from the stochastic variation in
protein levels between different cells.

Bistability has been observed in two-component systems. The conditions for
bistability in these systems have been discussed in \cite{igoshin08}.
This dynamical property has also been studied in \cite{amin13}. The case
treated there is that of a split kinase. This means that instead of one
kinase HK phosphorylating itself one kinase binds to a second which then
phosphorylates the first. These authors investigated bistability in different
models using both dynamical simulations and the Chemical Reaction Network
Toolbox. The latter is a computer programme which provides positive or negative
results on bistability on the basis of chemical reaction network theory.

Some signaling pathways include phosphorelays in which the phosphate group is 
transferred from one species to the next. They also have a cascade structure, 
in many cases with four layers. In some layers, a phosphate group can 
easily be lost by hydrolysis. Furthermore some species can be bifunctional, 
i.e. able to give as well as take a phosphate group to/from the species on the 
next layer. As for the MAPK cascade, such phosphorelays can be embedded in 
transcriptional feedback loops. See for example \cite{KFWCS} where 
mathematical results have been obtained on the form of response functions
and bistability in certain cases. There are many possible topologies for 
these systems and questions arise which are similar to those which could
be posed for the MAPK cascade: how complicated should the architecture of the 
network be in order to achieve a `good switch' property of the input-output 
relation? We saw that bistability could give an answer. Furthermore the 
methods explained in the previous sections could give insight on the dynamics 
beyond the steady states (heteroclinic structure, oscillations). 

\section{Summary and outlook}

The MAPK cascade is a system of chemical reactions occurring as a part of
many signal transduction networks. This cascade on its own has the potential 
to give rise to complicated dynamical behaviour such as multistability,
sustained oscillations and even chaos. The positive and negative feedback 
loops in which the cascade is embedded in real biological systems present even 
more possibilities for generating this type of phenomena. One way of trying
to obtain deeper insights into the conditions leading to different types of
dynamical behaviour is to carry out mathematically rigorous investigations
of systems of ordinary differential equations modelling the cascade. In
this context it is natural to start by studying small building blocks in
detail and build up from there. In particular we can pass from single layers
of the cascade to the full cascade without feedback and then the full
cascade with feedback.

In the first four sections of this paper the results on the MAPK cascade or parts of
it which have been proved rigorously are summarized. In the case of a 
single phosphorylation loop it could be shown that the dynamics are simple:
there is only one steady state and it is globally stable. For a series
of several phosphorylation loops it has been known for some time that 
multiple stationary solutions are possible and information is available on
their number. More recently it could be proved that for a system with two 
loops there are parameter values which give bistability, thus confirming
previous conclusions based on numerical simulations and heuristic
considerations. It could also be proved that for the MAPK cascade (or even
for the first two layers of it) there exist periodic solutions.  An
introduction is given to some of the mathematical techniques used to 
obtain these results, geometric singular perturbation theory and 
bifurcation theory. 

Cases are pointed out where further progress would be desirable. It could 
not yet be proved that the oscillations in the MAPK cascade are stable 
although simulations indicate that this is the case. In fact if they were 
not stable then it would be hard to find them by simulations. There are
also no analytical results available on the presence of chaos. Obtaining
results of this kind would require analysing bifurcations much more 
complicated than those treated up to now. It would also be valuable to 
obtain more results which in addition to showing that certain types of
behaviour can occur in a given system also give useful information on the
range of parameters for which it occurs. 
 
The fifth section contains some remarks of the influence of feedback 
loops on the dynamics of the cascade. This should be a fruitful field of
application for the techniques already developed for the cascade on its
own. Interestingly it seems based on simulations that the range of parameters 
for which bistability or sustained oscillations occur is increased by the 
presence of feedback loops and this may be important for the question of
whether these features are present for biologically reasonable 
parameters and if so, whether the ranges of these parameters are large
enough to allow the oscillations to be observed experimentally. 
 
The sixth section discusses some generalizations to other types of 
phosphorylation systems. The focus in this paper is on distributive 
and sequential phosphorylation and this mirrors a more general tendency in 
the theoretical literature on this subject. Replacing distributive by 
processive phosphorylation in some reactions appears to lead to a 
simplification of the dynamics. On the other hand distributive unordered
phosphorylation can lead to oscillations not present in the 
corresponding sequential case. Here again there is a lot of potential
for further analytical investigations. Remarks are also made on
the relations to signal transduction networks based on two-component
systems.  

Phosphorylation systems involved in signal transduction give rise to many
challenges for mathematical analysis which have only started to be addressed.
It is to be hoped that further progress in this direction will be rewarded
by a deeper understanding of the mechanisms of the biological processes
involved.

\end{document}